\documentclass[cameraready]{Interspeech}

\title{Analyzing Language and Geographical Variation in Speech Representations Across 60 Indic Languages}

\author[affiliation={1}]{Pavan Kumar}{J}
\author[affiliation={1}]{Agneedh}{Basu}
\author[affiliation={1}]{Pranav}{Bhat}
\author[affiliation={1}]{Sujith}{Pulikodan}
\author[affiliation={1}]{Visruth}{Sanka}
\author[affiliation={1}]{Nihar}{Desai}
\author[affiliation={2}]{Prasanta K}{Ghosh}

\address{
    $^1$ AI \& Robotics Technology Park (ARTPARK), I-Hub @ IISc, Bangalore, India \\
    $^2$Department of Electrical Engineering, Indian Institute of Science, Bangalore, India
}
\email{pavanjk@artpark.in}

\keywords{speech embeddings, multilingual speech processing, language identification}

\usepackage{comment}

\usepackage{graphicx}
\usepackage{subcaption}
\usepackage{multirow}
\usepackage{booktabs}
\usepackage{amsmath}
\usepackage{tikz}

\begin{document}

\maketitle

\begin{abstract}

Self-supervised speech encoders are often fine-tuned with language supervision, which can overlook geographical variation. To understand the learned representations under joint supervision of language and district compared to language-only supervision, we fine-tune Whisper-base and Wav2Vec2.0-base for classification tasks with joint language-district (386 classes) and language-only classification (60 languages). The language-district supervision improves district discrimination conditioned on language in the embedding space while strong marginal language classification. We analyze the structure of the learned embeddings using Normalized Conditional Mutual Information (NCMI), showing that language-district supervision produces global language clusters with structured within language subclusters aligned to district variation, enhancing geographical separability without degrading language-level organization.

\end{abstract}

\section{Introduction}

Speech embedding spaces learned by deep neural networks have become foundational to modern speech processing systems, supporting tasks such as language identification (LID), automatic speech recognition (ASR), speaker verification, and accent classification. Beyond downstream performance, recent work shows that supervised and self-supervised models learn embeddings that capture structured acoustic and linguistic information. Törö et al.~\cite{toro2025neighbors} demonstrate that distances in multilingual speech embedding spaces reflect typological and phonological relationships across languages, suggesting that models implicitly organize languages according to linguistic similarity. Likewise, Bartelds and Wieling~\cite{bartelds2022quantifying} show that neural acoustic representations encode fine-grained pronunciation variation, enabling direct quantification of language variation from speech. Together, these findings indicate that learned speech embeddings preserve meaningful acoustic–linguistic structure.

Building on this representational capacity, research on Indic language identification (LID) spans supervised, self-supervised, and cross-corpora approaches. CNN and ANN models trained on log-Mel spectrograms have been widely applied to languages~\cite{mukherjee2019deep, godbole2020indian}. Cross-corpora robustness has been improved using MFCC features with TDNN and ECAPA-TDNN architectures, often combined with normalization and enhancement techniques including CMS, CMVN, RASTA, and PCEN~\cite{dey2021cross, dey2023cross}. Recent supervised LID work increasingly leverages self-supervised pre-training. Studies showed that phonotactic representations learned from unlabeled speech capture language-discriminative structure effectively~\cite{ramesh2021self}. Adaptive multilingual pre-training further improves fine-grained language and dialect identification under low-resource settings~\cite{shaik2023self}. 

Unified ASR-LID frameworks integrate language identification within encoder-decoder architectures for Indic languages~\cite{jakhar2024unified}, while domain-adapted Whisper-base variants demonstrate the effectiveness of hybrid multilingual and multitask training for spoken LID and transcription~\cite{11377770}. Collectively, these approaches show that supervised and semi-supervised learning effectively encode language identity in multilingual speech representations. Beyond language-level discrimination, embedding geometry has been shown to capture dialectal and geographical structure. Dialect-based supervision shapes representation spaces, producing systematic but sometimes unstable clustering~\cite{dunn2023variation}. Accent and dialect information can also be reliably extracted from pre-trained speech and speaker embeddings~\cite{ghorbani2024advanced}. In the Indian context, jointly modeling dialect identification and ASR improves downstream performance, highlighting the importance of geographical variation in multilingual systems~\cite{kumar2025jointly, kumar2025improving}. These findings suggest that district or dialect-level variation constitutes structured information within speech representations rather than noise.

Existing studies on Indic language identification often balance between broad multilingual coverage and fine-grained geographical or dialect modeling. Some works focus on a smaller set of languages to enable detailed analysis of dialectal or regional variation, while others scale to larger language inventories with supervision primarily at the language level. As a result, the interaction between supervision granularity and the internal geometry of multilingual speech embeddings particularly in capturing systematic within language geographical variation at scale remains underexplored.


In this work, we address this gap by analyzing multilingual speech representations across 60 Indic languages spanning 165 districts, resulting in 386 language–district classes with balanced per-class data. We fine-tune Whisper-base~\cite{radford2023robust} and Wav2Vec~2.0-base~\cite{baevski2020wav2vec} under two objectives: (i) language-only classification and (ii) joint language–district classification. We evaluate representations using classification accuracy, language-conditioned probing with logistic regression (reporting both accuracy and F1 score), and a metric computed on the embedding space, Normalized Conditional Mutual Information (NCMI), to quantify the structure of the learned embeddings. Our results show that fine-grained supervision enhances district-level discriminability while preserving strong language separability and induces structured within-language subclusters in the embedding space. These findings demonstrate that supervision granularity systematically shapes multilingual speech representation geometry in linguistically diverse settings. To our knowledge, no prior work has explored joint language-district classification at this scale for Indian languages.

\section{Dataset}

We conduct experiments on a subset of the \textit{Vaani} speech corpus~\cite{vaani2025}, a large-scale Indian speech dataset with fine-grained geographical metadata, making it well suited for analyzing the interaction between language-level supervision and geographical variation in multilingual speech models.

We select 60 languages spanning 165 districts, yielding 386 language-district classes. This subset balances typological and geographical coverage while ensuring sufficient data per class. For the joint language-district setup, each class contains approximately 3 hours of speech, totaling 1158 hours. Data are split at the speaker level into training (2.4 h), validation (0.2 h), and test (0.4 h) partitions to prevent speaker overlap.

In addition to the joint setup, we construct two language-only regimes. The balanced setting contains 3 hours per language (180 hours total), uniformly sampled across districts to preserve intra-language geographical diversity. The unbalanced setting uses the same 1158 hours as the joint setup but collapses labels to language only, retaining natural data imbalance. This design isolates the effect of supervision granularity while controlling for data scale and speaker independence.

\begin{table}[!h]
\centering
\small
\caption{Classification performance (Accuracy / F1(mean), \%).}
\label{tab:marginal_acc}

\begin{tabular*}{\columnwidth}{@{\extracolsep{\fill}}lcccccc}
\toprule
& \multicolumn{2}{c}{Overall} 
& \multicolumn{2}{c}{Lang} 
& \multicolumn{2}{c}{Distict} \\
\cmidrule(lr){2-3} \cmidrule(lr){4-5} \cmidrule(lr){6-7}
\textbf{Setup} 
& Acc & F1 
& Acc & F1 
& Acc & F1 \\
\midrule
\multicolumn{7}{l}{\textbf{Whisper-base}} \\
LD-386 & 44.27 & - & \textbf{84.79} & \textbf{79.00} & 47.09 & 43.16 \\
L-60 & 66.99 & 67.13 & 66.99 & 67.13 & - & - \\
L-60-FD & 84.77 & 78.10 & 84.77 & 78.10 & - & - \\
\midrule
\multicolumn{7}{l}{\textbf{Wav2Vec2.0-base}} \\
LD-386 & 21.73 & - & 71.56 & 55.66 & 23.91 & 21.87 \\
L-60 & 64.16 & 66.43 & 64.16 & 66.43 & - & - \\
L-60-FD & 81.93 & 73.61 & \textbf{81.93} & \textbf{73.61} & - & - \\
\bottomrule
\end{tabular*}
\end{table}



\begin{table}[t]
\centering
\caption{Language-conditioned probing for district classification performance using embeddings from different training regimes. Accuracy (\%) and mean F1 are shown for Whisper-base and Wav2Vec2.0-base}
\label{tab:lc_district_perf}
\begin{tabular*}{\columnwidth}{@{\extracolsep{\fill}}l cc cc}
\hline
\textbf{Setup} & \multicolumn{2}{c}{\textbf{Whisper-base}} & \multicolumn{2}{c}{\textbf{Wav2Vec2.0-base}} \\
 & Acc (\%) & F1 (mean) & Acc (\%) & F1 (mean) \\
\hline
LD-386      & \textbf{91.59} & \textbf{91.29} & \textbf{87.56} & \textbf{87.14} \\
L-60        & 81.41 & 80.89 & 77.72 & 77.05 \\
L-60-FD     & 86.88 & 86.46 & 59.92 & 58.53 \\
\hline
\end{tabular*}
\end{table}

\section{Experimental Setup \& Embedding Analysis}
\label{sec:model}

\begin{table*}[t]
\centering
\small
\caption{Language-wise marginal F1 scores. Languages are columns; setups are grouped by model.}
\resizebox{\textwidth}{!}{
\begin{tabular}{lcccccccccccccccccc}
\toprule
Setup & Awadhi & Bearybashe & Garhwali & Gondi & Halbi & Haryanvi & Hindi & IduMishmi & Karbi & Kashmiri & Lotha & Nimadi & Punjabi & Rajasthani & Rengma & Sadri & Surjapuri & Tulu \\
\midrule
\multicolumn{19}{l}{\textbf{Whisper-base}} \\
LD-386 & 91.1 & 49.23 & \textbf{68.37} & \textbf{90.4} & 79.86 & \textbf{66.53} & \textbf{87.27} & 82.3 & 84.87 & 84.99 & \textbf{87.66} & \textbf{84.01} & 86.76 & 72.38 & \textbf{91.87} & \textbf{51.91} & \textbf{65.54 }& 71.15 \\
L-60 & 72.65 & \textbf{53.73} & 43.55 & 76.73 & 71.72 & 35.56 & 60.77 & 79.47 & 80.41 & 60.04 & 84.99 & 56.66 & 64.92 & 49.36 & 88.94 & 24.35 & 42.96 & 65.6 \\
L-60-FD & \textbf{95.01} & 46.43 & 65.14 & 87.2 & \textbf{81.41} & 65.35 & 87.02 & \textbf{86.3} & \textbf{91.49} & \textbf{88.27} & 86.21 & 81.71 & \textbf{87.64} & \textbf{72.7}5 & 91.65 & 38.16 & 65.1 & \textbf{72.73} \\
\midrule
\multicolumn{19}{l}{\textbf{Wav2Vec2.0-base}} \\
LD-386 & 84.66 & 4.42 & 29.17 & 59.28 & 45.76 & 43.12 & 81.51 & 34.36 & 56.47 & 49.45 & 31.69 & 37.76 & 68.0 & 53.28 & 44.37 & 21.94 & 52.28 & 20.21 \\
L-60 & 82.37 & 48.18 & 48.33 & 83.8 & 68.74 & 51.97 & 54.69 & 77.24 & 85.15 & 72.73 & \textbf{76.31} & 44.52 & 71.55 & 56.09 & 85.32 & 23.91 & 57.76 & 58.01 \\
L-60-FD & \textbf{89.98} & \textbf{56.08} & \textbf{60.99} & \textbf{91.04} & \textbf{76.76} & \textbf{52.59} & \textbf{84.41} & \textbf{77.79} & \textbf{86.67} & \textbf{74.91} & 74.21 & \textbf{72.87} & \textbf{89.68} & \textbf{67.82} & \textbf{86.12} & \textbf{38.6} & \textbf{64.07} & \textbf{67.72} \\
\bottomrule
\end{tabular}
}
\label{tab:language_f1_wide_split}
\end{table*}

\begin{table*}[t]
\centering
\small
\caption{Language-conditioned district classification performance (F1-score) using multinomial logistic regression, evaluated within language-specific embedding subspaces.}
\resizebox{\textwidth}{!}{
\begin{tabular}{lcccccccccccccccccc}
\toprule
Setup & Assamese & Bengali & Bhojpuri & Chhattisgarhi & English & Gujarati & Hindi & Kannada & Maithili & Malayalam & Marathi & Marwadi & Nepali & Oriya & Punjabi & Tamil & Telugu & Urdu \\
\midrule
\multicolumn{19}{l}{\textbf{Whisper-base}} \\
LD-386 & \textbf{90.02} & \textbf{85.12} & \textbf{92.7} & \textbf{92.36} & \textbf{96.18} & \textbf{98.21} & \textbf{64.08} & \textbf{68.37} & \textbf{90.75} & \textbf{83.68} & \textbf{80.51} & \textbf{94.21} & \textbf{88.85} & \textbf{89.6} & \textbf{98.19} & \textbf{80.16} & \textbf{72.55} & \textbf{98.41} \\
L-60 & 75.78 & 62.81 & 76.07 & 73.93 & 82.62 & 90.12 & 38.73 & 50.13 & 72.68 & 65.14 & 60.54 & 80.59 & 75.42 & 77.36 & 89.71 & 63.8 & 47.23 & 90.39 \\
L-60-FD & 80.12 & 73.62 & 86.15 & 86.17 & 89.53 & 93.27 & 50.84 & 58.53 & 85.42 & 74.4 & 68.44 & 89.58 & 80.62 & 80.34 & 95.87 & 71.25 & 60.04 & 96.07 \\
\midrule
\multicolumn{19}{l}{\textbf{Wav2Vec2.0-base}} \\
LD-386 & \textbf{83.52} & \textbf{77.04} & \textbf{86.4} & \textbf{85.72} & \textbf{89.56} & \textbf{92.36} & \textbf{57.8} & \textbf{61.47} & \textbf{86.67} & \textbf{72.96} & \textbf{70.82} & \textbf{89.95} & \textbf{80.62} & \textbf{83.95} & \textbf{95.86} & \textbf{75.21} & \textbf{62.88} & \textbf{94.61} \\
L-60 & 59.87 & 50.64 & 74.67 & 73.86 & 64.01 & 83.75 & 41.77 & 46.72 & 76.94 & 56.72 & 55.13 & 81.55 & 68.08 & 74.04 & 85.6 & 59.53 & 41.3 & 87.27 \\
L-60-FD & 52.91 & 20.05 & 42.4 & 45.27 & 30.49 & 60.18 & 14.79 & 22.47 & 52.58 & 31.59 & 26.81 & 62.86 & 55.02 & 50.4 & 56.71 & 35.39 & 12.88 & 59.78 \\
\bottomrule
\end{tabular}
}
\label{tab:language_conditioned_district_wide_split}
\end{table*}

\subsection{Pretrained Models Architecture}

We evaluate two pretrained speech encoders: Whisper-base and Wav2Vec2.0-base. Whisper-base operates on log-Mel spectrogram inputs, while Wav2Vec2.0-base processes raw waveforms via a convolutional feature encoder followed by a Transformer context network. In both cases, the encoder produces frame-level representations of dimension $D$, and all layers are fine-tuned end-to-end without freezing, allowing adaptation to different supervision objectives.

We analyze Wav2Vec2.0-base and Whisper-base due to their complementary properties. Wav2Vec2.0-base, pre-trained in a self-supervised manner on large English corpora including LibriSpeech \cite{panayotov2015librispeech} provides high-quality embeddings to examine how an English-trained model captures fine-grained acoustic and language-specific structure in unseen Indic languages. Whisper-base, trained on large-scale multilingual speech, serves as a benchmark for cross-lingual representation. Using these architectures allows us to isolate the effects of model capacity and pretraining regime on embedding structure without confounding domain-specific biases.

\subsection{Utterance-Level Embedding Representations}

For speech utterance a frame-level output obtained from the encoder of either \textit{Whisper-base} or \textit{Wav2Vec2.0-base}, are converted to utterance-level embeddings using attention-based temporal pooling. For Whisper-base, $D=512$, while for Wav2Vec2.0-base, $D=768$. Given frame-level encoder outputs 
$\mathbf{X} \in \mathbb{R}^{T \times D}$ 
(with $T$ frames and embedding dimension $D$), 
attention-based temporal pooling computes the utterance embedding as

\[
\mathbf{h}_\text{utt} 
= \mathbf{X}^\top \boldsymbol{\alpha}, 
\quad 
\boldsymbol{\alpha} = \text{softmax}\left(f_\text{attn}(\mathbf{X})\right),
\]

where 
$f_\text{attn} : \mathbb{R}^{T \times D} \rightarrow \mathbb{R}^{T}$ 
is a two-layer feed-forward network with $\tanh$ activation that produces a scalar score for each frame, 
$\boldsymbol{\alpha} \in \mathbb{R}^{T}$ are the normalized attention weights over time, 
and $\mathbf{h}_\text{utt} \in \mathbb{R}^{D}$ is the resulting fixed-dimensional utterance embedding (512 for Whisper-base and 768 for Wav2Vec2.0-base). 

This formulation enables the model to assign higher weights to acoustically salient frames when forming compact utterance-level Embedding representations.

\subsection{Normalized Conditional Mutual Information (NCMI)}

To quantify how well embeddings capture district-level structure while respecting language, we compute \textit{Normalized Conditional Mutual Information (NCMI)} on the test embeddings. Let the test dataset be $\mathcal{D}_{\text{test}} = \{ (x_i, \ell_i, d_i) \}_{i=1}^{N},$

where $x_i \in \mathbb{R}^D$ is the embedding of sample $i$ extracted from the attention pooling layer of the fine-tuned model, $\ell_i$ is its ground-truth language, and $d_i$ is its ground-truth district. For each sample $i$, we find its $k$ nearest neighbors in the embedding space, denoted $\mathcal{N}_k(i)$. We then restrict to neighbors that share the same language $\ell_i$. Let $p_i(d_i, n \mid \ell_i)$ be the fraction of these neighbors having district label $n$, jointly with the sample’s district $d_i$. The per-sample conditional mutual information between the sample’s district and its neighbors, conditioned on language, is
\vspace{-0.2cm}
\[
I_i(D; \mathcal{N}_k \mid L) = \sum_{n \in \mathcal{N}_k(i)} 
p_i(d_i, n \mid \ell_i) \, \log \frac{p_i(d_i, n \mid \ell_i)}{p_i(d_i \mid \ell_i) \, p_i(n \mid \ell_i)},
\]
\vspace{-0.2cm}
\[
H_i(D \mid L) = \sum_{n \in \mathcal{N}_k(i)} p_i(n \mid \ell_i) \, \log \frac{1}{p_i(n \mid \ell_i)},
\]

where $p_i(d_i \mid \ell_i)$ and $p_i(n \mid \ell_i)$ are the marginal probabilities among neighbors with language $\ell_i$. The per-sample conditional entropy of district labels given language is $H_i(D \mid L)$, which measures the uncertainty of the neighbor districts conditioned on the sample’s language. Finally, the dataset-level normalized conditional mutual information is obtained by averaging over all $N$ samples:

\[
\text{NCMI}_k = \frac{1}{N} \sum_{i=1}^{N} \frac{I_i(D; \mathcal{N}_k \mid L)}{H_i(D \mid L)}.
\]

High NCMI indicates that embeddings are organized such that districts form coherent local clusters within each language, and its variation with $k$ captures the hierarchical structure of the embedding space. This formulation mirrors the implementation in our Python code, where neighbors, joint probabilities, and weighting are explicitly computed from the embeddings. While we present NCMI conditioned on language, the same formulation can be applied by conditioning on district instead, allowing evaluation of how well embeddings capture language discriminative structure within geographic regions. 
To quantify how conditional structure persists across neighborhood scales, we integrate the NCMI curve in $\log k$, thus capturing the cumulative strength of conditional information from local to global neighborhoods. We then compute a normalized asymmetry score $\Delta_{\text{scale}}$ as the relative difference between district-conditioned and language-conditioned AUC . This scalar provides a scale-invariant measure of conditional dominance: negative values indicate stronger persistence of language-level structure, while values closer to zero reflect more balanced multi-scale organization.

The implementation of our experiments utilizes \texttt{PyTorch}~\cite{paszke2019pytorch} to build the attention pooling layer, \texttt{HuggingFace Transformers}~\cite{wolf2020transformers} for pretrained models and fine-tuning, and \texttt{scikit-learn}~\cite{pedregosa2011scikit}, \texttt{SciPy}~\cite{virtanen2020scipy}, and \texttt{NumPy}~\cite{harris2020array} for embedding processing, calculating NCMI, and statistical computations.

\subsection{Fine-Tuned Model Architecture}

From Utterance-Level Embedding $\mathbf{h_{utt}} \in \mathbb{R}^{D}$ We apply a feed-forward classifier with one hidden layer (ReLU + dropout) followed by a linear projection. The logits optimized using cross-entropy loss. We consider three supervision settings: \textbf{L-60}, language-only classification over 60 languages; \textbf{LD-386}, joint language-district classification over 386 classes; and \textbf{L-60-FD}, language-only classification over the full dataset used in LD-386 (district labels collapsed to 60 languages). 



\subsection{Experimental Setup}

All models are fine-tuned for classification and overall performance is reported on a held out test set. For LD-386, marginal language probabilities are computed by summing probabilities across districts belonging to the same language, enabling direct comparison with L-60 and L-60-FD. To analyze embedding structure, we first perform linear probing by training a logistic regression classifier for district prediction conditioned on language using the learned utterance-level Embedding $\mathbf{h_{utt}}$ across all setups. We further compute Normalized Conditional Mutual Information (NCMI) on utterance-level Embedding $\mathbf{h_{utt}}$ under two conditions: conditioning on language, $I(D_i;D_j \mid L)$, and conditioning on district, $I(L_i;L_j \mid D)$. 
NCMI is estimated using $k$-nearest neighbors with Euclidean distance for varying $k$ to study local and global geometric structure. All experiments are trained using the AdamW optimizer on a single NVIDIA L4 GPU (24\,GB).






\begin{figure*}[t]
    \centering

    \begin{subfigure}[t]{0.24\textwidth}
        \centering
        \includegraphics[width=\linewidth]{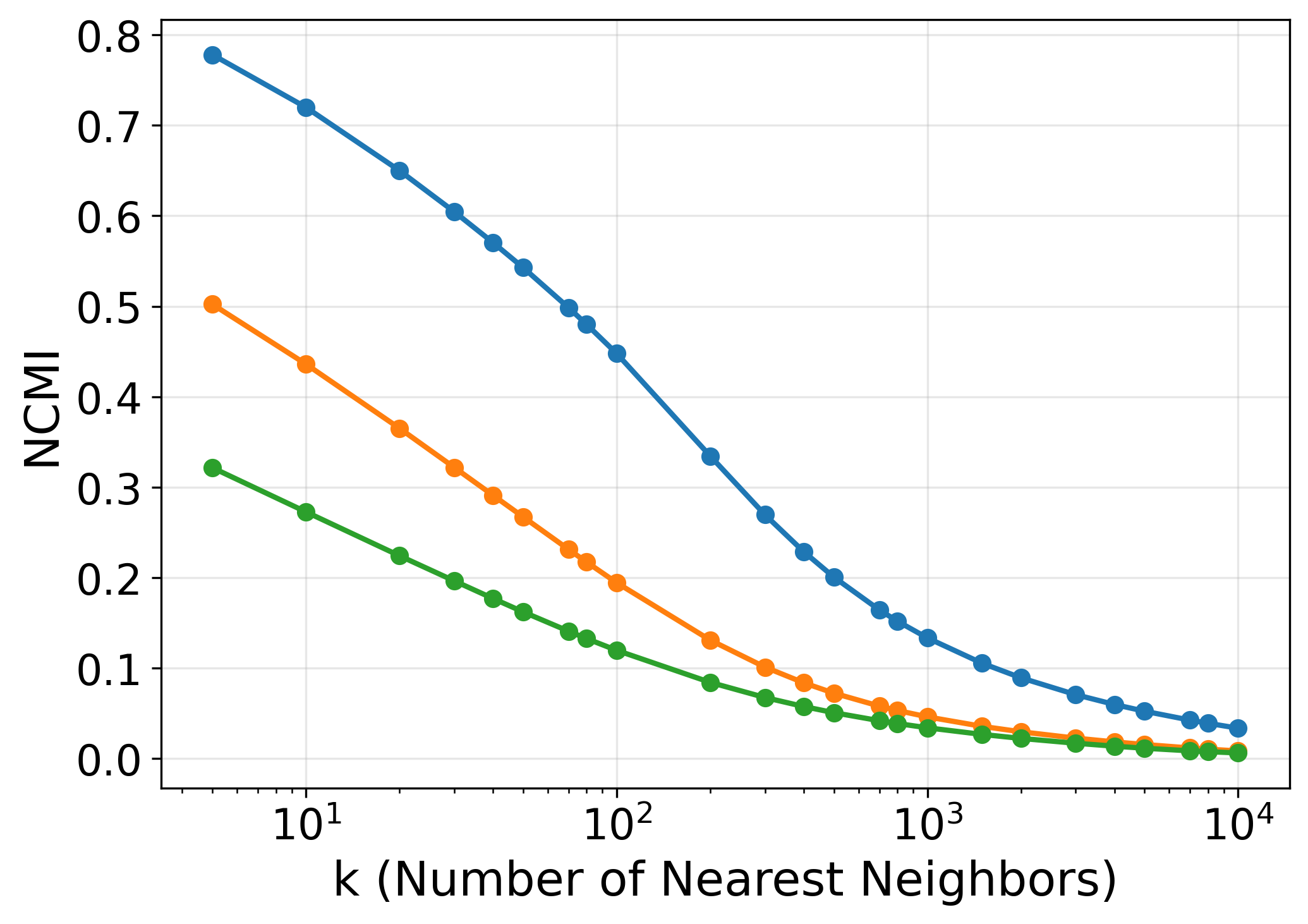}
        \caption{Whisper-base (Lang-Cond.)}
        \label{fig:whisper_lang_ncmi}
    \end{subfigure}
    \hfill
    \begin{subfigure}[t]{0.24\textwidth}
        \centering
        \includegraphics[width=\linewidth]{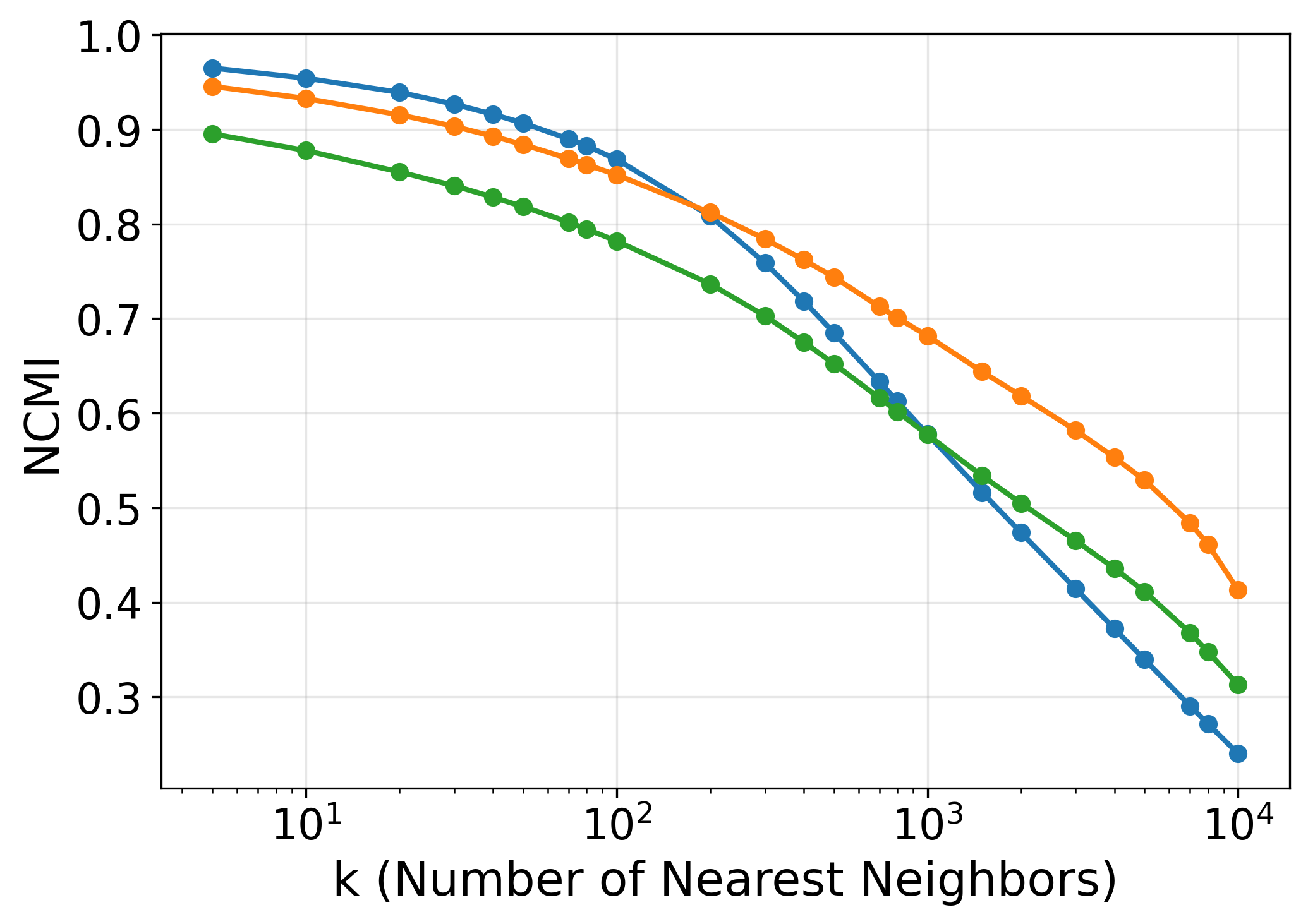}
        \caption{Whisper-base (Dist-Cond.)}
        \label{fig:whisper_dist_ncmi}
    \end{subfigure}
    \hfill
    \begin{subfigure}[t]{0.24\textwidth}
        \centering
        \includegraphics[width=\linewidth]{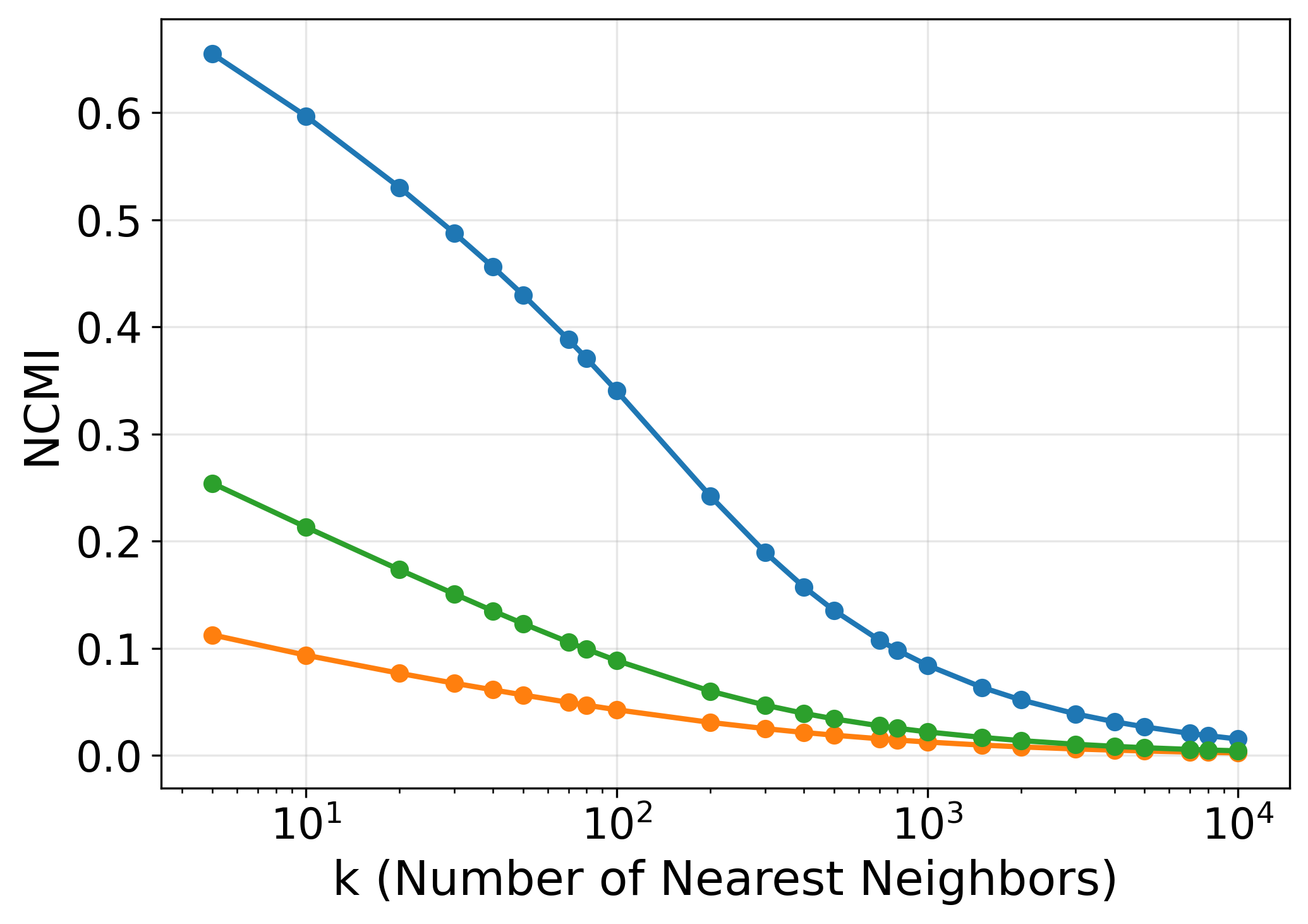}
        \caption{Wav2Vec2.0-base (Lang-Cond.)}
        \label{fig:wav2vec2_lang_ncmi}
    \end{subfigure}
    \hfill
    \begin{subfigure}[t]{0.24\textwidth}
        \centering
        \includegraphics[width=\linewidth]{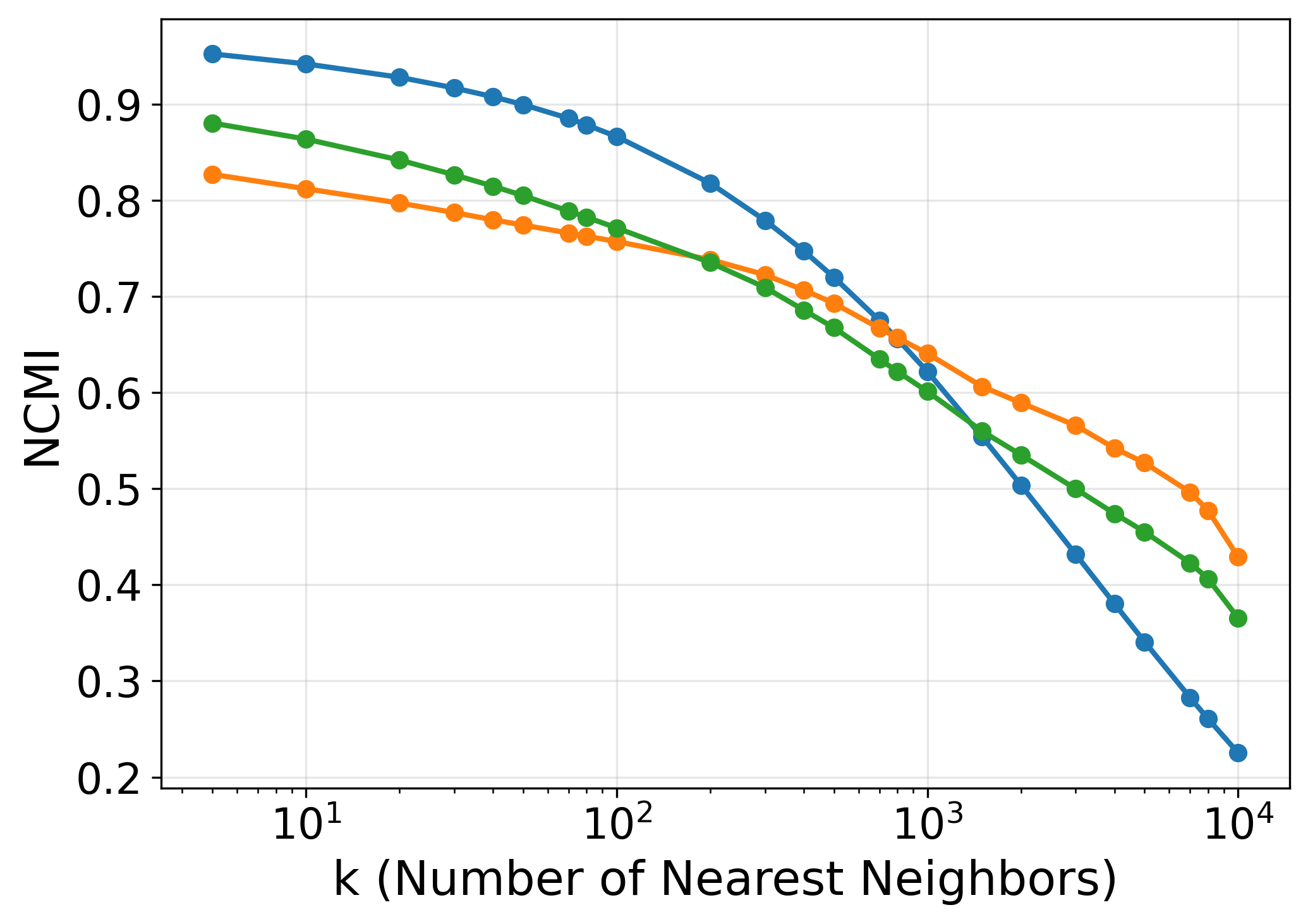}
        \caption{Wav2Vec2.0-base (Dist-Cond.)}
        \label{fig:wav2vec2_dist_ncmi}
    \end{subfigure}


    \vspace{-0.2cm}
    \begin{center}
    \footnotesize
    \textbf{Experiment Setup:}
    \tikz[baseline=-0.5ex]{
        \draw[blue, thick] (0,0) -- (1cm,0);
        \fill[blue] (0.5cm,0) circle[radius=1.5pt];
    } LD-386 \quad
    \tikz[baseline=-0.5ex]{
        \draw[orange, thick] (0,0) -- (1cm,0);
        \fill[orange] (0.5cm,0) circle[radius=1.5pt];
    } L-60-FD \quad
    \tikz[baseline=-0.5ex]{
        \draw[green!60!black, thick] (0,0) -- (1cm,0);
        \fill[green!60!black] (0.5cm,0) circle[radius=1.5pt];
    } L-60
    \end{center}
    \vspace{-0.4cm}

    \caption{Normalized Conditional Mutual Information (NCMI) as a function of $k$. 
    From left to right: Whisper-base language-conditioned, Whisper-base district-conditioned, 
    Wav2Vec2.0-base language-conditioned, and Wav2Vec2.0-base district-conditioned.}
    \label{fig:ncmi_comparison_single_row}
\end{figure*}




\section{Experiments and Results}



\subsection{Language Classification Performanc}

Table~\ref{tab:marginal_acc} reports overall classification accuracy, marginal language performance (obtained by aggregating district probabilities for LD-386), and district-level performance for joint models. 
For Whisper-base, LD-386 achieves strong marginal language performance (84.79\% accuracy, 79.00 F1), comparable to L-60-FD (84.77\%, 78.10 F1), while additionally enabling district prediction (47.09\% accuracy, 43.16 F1). 
For Wav2Vec2.0-base, L-60-FD attains the highest language accuracy (81.93\%), exceeding LD-386 (71.56\%), though LD-386 uniquely captures district-level structure (23.91\% accuracy, 21.87 F1). 
Thus, joint supervision preserves competitive language separability while organizing district information that language-only models cannot directly model.

Table~\ref{tab:language_f1_wide_split} further presents language-wise marginal F1 scores. 
For Whisper-base, LD-386 is competitive or superior across many languages (e.g., Garhwali, Gondi, Haryanvi, Lotha, Nimadi), indicating that explicit geographical supervision enhances robustness without harming language discrimination. 
For Wav2Vec2.0-base, increasing data scale in L-60-FD often yields the highest marginal F1, though several languages remain comparable under LD-386. Due to space constraints, only a representative subset of languages is shown; the remaining languages follow trends consistent with the aggregate results. Overall, joint language-district supervision maintains strong marginal language performance, whereas language-only supervision primarily benefits from increased data scale.

\subsection{Language-Conditioned Probing and Embedding Structure}

We evaluate district structure using language-conditioned probing with a multinomial logistic regression classifier trained independently within each language-specific embedding subspace. 
Table~\ref{tab:lc_district_perf} reports mean F1 scores. Joint language-district supervision (LD-386) achieves the strongest district performance for both models. 
For Whisper-base, LD-386 reaches 91.29 mean F1, outperforming L-60 (80.89) and L-60-FD (86.46). 
For Wav2Vec2.0-base, LD-386 attains 87.14 mean F1, compared to 77.05 for L-60, while performance drops sharply to 58.53 under L-60-FD. 
The substantial degradation in Wav2Vec2.0-base when scaling language-only training indicates collapse of intra-language district structure.

The same trend appears at the per-language level Table~\ref{tab:language_conditioned_district_wide_split}. 
For example, under Whisper-base, LD-386 improves Hindi and Kannada district F1 by large margins over L-60. 
For Wav2Vec2.0-base, L-60-FD shows severe degradation across several languages (e.g., Hindi and Telugu), whereas LD-386 remains consistently strong. 
These representative cases reflect the broader pattern across languages. Combined with the marginal language results from the previous section, these findings show that explicit district supervision organizes fine-grained geographical variation in a linearly accessible manner. Language-only supervision particularly for Wav2Vec2.0-base tends to collapse intra-language structure as data scale increases, while joint supervision yields more structured and informative embeddings without sacrificing language separability.

\subsection{Embedding Structure via NCMI}

Figure~\ref{fig:ncmi_comparison_single_row} presents NCMI as a function of neighborhood size $k$ (log-scale). 
Smaller $k$ captures highly local embedding neighborhoods, while larger $k$ reflects increasingly global structure. 
Evaluation over $N=114{,}596$ test samples produces smooth and stable trends. Under language conditioning (Fig.~\ref{fig:whisper_lang_ncmi}, Fig.~\ref{fig:wav2vec2_lang_ncmi}), LD-386 consistently achieves higher district-level NCMI across nearly all neighborhood sizes for both Whisper-base and Wav2Vec2.0-base. 
The separation is most pronounced at small $k$, indicating stronger preservation of fine-grained district structure within language clusters. Although NCMI decreases with increasing $k$ as neighborhoods become more global, the relative ordering remains stable. The effect is particularly sharp for Wav2Vec2.0-base, where language-only supervision shows clearer intra-language collapse, Whisper-base exhibits comparatively more stable district retention across scales.

Under district conditioning (Fig.~\ref{fig:whisper_dist_ncmi}, Fig.~\ref{fig:wav2vec2_dist_ncmi}), all models exhibit high NCMI at small $k$, confirming strong local language purity. 
As $k$ increases, NCMI gradually declines. 
For Wav2Vec2.0-base, language-only training produces stronger global language clustering at large $k$, while LD-386 maintains more balanced behavior. 
In contrast, Whisper-base shows more stable trends, with LD-386 preserving district structure without sacrificing language separability. 
Consistent with earlier marginal classification results, LD-386 achieves comparable or better language accuracy, indicating that its lower large-scale NCMI does not imply weaker language discrimination.

To summarize, we report the normalized AUC asymmetry score for the LD-386 setup. 
For Whisper-base, $\mathrm{AUC}_{D|L}=2.68$ and $\mathrm{AUC}_{L|D}=5.43$, yielding $\Delta_{\text{scale}}=-0.34$. 
For Wav2Vec2.0-base, $\mathrm{AUC}_{D|L}=2.08$ and $\mathrm{AUC}_{L|D}=5.43$, yielding $\Delta_{\text{scale}}=-0.45$. 
The negative values indicate stronger persistence of language-level structure across neighborhood scales. 
The larger magnitude for Wav2Vec2.0-base reflects greater coarse-level dominance, whereas Whisper-base retains relatively more district-level structure, resulting in a more balanced embedding geometry.

The asymmetric NCMI behavior supports a hierarchically organized embedding space, district coherence is strongest locally, while language separation remains stable globally. 
Importantly, no explicit hierarchical constraint was imposed during training the structure emerges implicitly from joint language-district supervision. Across experiments, Whisper-base exhibits more stable and stronger NCMI trends than Wav2Vec2.0-base, likely reflecting pretraining alignment, as Whisper-base was trained on multilingual (including Indic) data whereas Wav2Vec2.0-base was primarily pretrained on English.

\section{Conclusion and Future Work}

We show that supervision granularity shapes multilingual speech embeddings: joint language-district training, while increasing task difficulty, preserves strong marginal language performance and markedly improves district discriminability. Language-conditioned probing and NCMI analysis reveal that fine-grained supervision forms structured within-language subclusters reflecting geographical variation, whereas language-only training yields region-invariant embeddings. Whisper-base encoders exhibit stronger and more stable geographical encoding than Wav2Vec2.0-base. Overall, fine-grained supervision enhances embedding structure without compromising language separability. 
Future work will explore leveraging this structure for low-resource language-district pairs via representation sharing and its impact on downstream tasks such as automatic speech recognition and text-to-speech in multilingual settings.

\bibliographystyle{IEEEtran}
\bibliography{mybib}

\end{document}